\documentclass[prl, showpacs, twocolumn, floatfix]{revtex4}

\usepackage{graphicx}
\usepackage{amsmath, amsfonts, amssymb, bm}
\usepackage[]{psfrag}

\psfragscanoff
\begin{document}

\title{ Two-center dielectronic recombination}
%Strongly enhanced recombination via two-center electronic correlations }

\author{C. M\"uller$^1$}
\author{A. B. Voitkiv$^1$}
\author{J. R. Crespo L\'opez-Urrutia$^1$}
\author{Z. Harman$^{1,2}$}
\affiliation{$^1$Max-Planck-Institut f\"ur Kernphysik, Saupfercheckweg 1, 69117 Heidelberg, Germany\\
$^2$ExtreMe Matter Institute (EMMI), Planckstrasse 1, 64291 Darmstadt, Germany}

\date{\today}

\begin{abstract}
In the presence of a neighboring atom, electron-ion recombination 
can proceed resonantly via excitation of an electron in the atom, 
with subsequent relaxation through radiative decay.
It is shown that this two-center dielectronic process can 
largely dominate over single-center radiative recombination at 
internuclear distances as large as several nanometers. 
The relevance of the predicted process is demonstrated 
by using examples of water-dissolved alkali 
cations and warm dense matter.
\end{abstract}
 

\maketitle

Recombination of free electrons with atomic or molecular ions is a fundamental process of general interest and relevance to various scientific disciplines \cite{Hahn,AMueller}. Recombination into single atomic centers may proceed in three ways: First, the electron can be captured into a bound state upon photoemission, referred to as radiative recombination (RR) being the time-inverse of photoionization. Second, for certain energies of the incident electron, the recombination can proceed resonantly via formation of an autoionizing state (time-reversed Auger decay); afterwards, the system stabilizes through radiative deexcitation. This dielectronic recombination (DR) process is particularly important for low-charged ions.
Finally, three-body recombination, in which an electron is captured by an ion transferring excess energy to another free electron, dominates in high-density plasmas.

When an atom is not isolated in space but close to another atom, the electronic structures at the two centers can be coupled by long-range electromagnetic interactions leading to a variety of interesting phenomena. For example, interatomic electron-electron correlations are responsible 
for various deexcitation processes in slow atomic collisions \cite{Smirnov}, including Penning ionization,  the population inversion in a He-Ne laser, and the energy transfer in quantum optical ensembles \cite{Ficek} or cold Ryd\-berg gases \cite{Weidemueller}. They also play an important role in biological systems as F\"orster resonances between chromophores \cite{Forster}. 
Another interesting realization of two-center electron-electron coupling  
is represented by a process in which the electronic excitation energy 
of one of the atoms cannot be quickly released through a forbidden 
(single-center) Auger decay and is instead transferred to the partner 
atom resulting in its ionization. Stimulated by detailed theoretical 
predictions \cite{ICD}, this interatomic decay process 
has been observed in recent years in various systems such as van der Waals clusters \cite{clusters}, rare gas dimers \cite{dimers}, and water molecules \cite{ICDexpH2O}.  

Very lately, a process has been calculated \cite{ICEC} 
where capture of an incident electron by an ion (or atom) proceeds 
via the Coulomb interaction of this electron with a neighboring atom 
leading to ionization of the latter. Keeping the total charge of the two centers 
unchanged, such a process effectively results in an interatomic electron exchange.

In this Letter, we introduce a process, in which  
an incident electron can be captured due to resonant electronic 
correlations involving two neighboring atomic centers.
The centers may be (but are not limited to) atoms, ions or molecules. 
In this process, which may be termed two-center 
dielectronic recombination (2CDR), the electron is captured by one of the centers  
with simultaneous resonant 
excitation of the other center which subsequently
deexcites via spontaneous radiative decay  (see Fig.1).
In contrast to the process considered in \cite{ICEC}, 
2CDR is a genuine capture process in which 
the total charge of the centers is changed.
Using examples from various fields of science, we shall show that 2CDR can very significantly contribute to recombination exceeding the usual single-center RR by orders of magnitude.

\begin{figure}[b]
\vspace{-0.3cm}
\begin{center}
\resizebox{6cm}{!}{\includegraphics{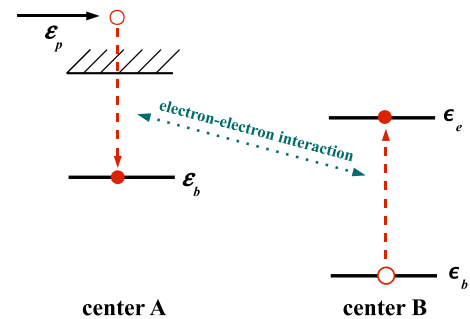}}
\vspace{-0.3cm}
\caption{Scheme of two-center dielectronic recombination (2CDR). Depicted is the first step, where an electron is captured at center $A$ with simultaneous excitation of an atom at center $B$. Afterwards, atom $B$ deexcites via photoemission.}
\end{center} 
\end{figure}

In order to emphasize the basic physics of 2CDR, let us
consider recombination in a simple system 
consisting of an incident electron, a bare nucleus 
(the nucleus $A$), an atom with one ``active'' electron 
(the atom $B$), and the radiation field. 
Both the bare and atomic nuclei are supposed to be at rest. 
We take the position of the bare nucleus as the origin 
and denote the coordinates 
of the atomic nucleus, the incident and atomic electrons 
by ${\bf R}$, ${\bf r}_1$ and ${\bf r}_2={\bf R} + \mbox{\boldmath{ $\xi$ }} $, 
respectively, where $\mbox{\boldmath{ $\xi$ }}$ 
is the position of the atomic electron with respect 
to the atomic nucleus.

We shall assume that the distance  
$ R $ is not too large, $ R \ll c/\omega_0$, where 
$c$ is the speed of light and $\omega_0$ 
the difference between the initial and final energies of the 
incident electron (its transition frequency). 
Then one can neglect the retardation effects and 
treat the interaction between the charged particles 
as instantaneous Coulombic. 
Besides, we shall also suppose that the distance $R$ 
is not too small either such that one can still 
speak about the individual 
subsystems: ``the incident electron + the nucleus $A$'' and 
the atom $B$. Under such conditions the energy exchange 
between the electrons is most efficient for dipole-allowed 
transitions and, restricting our attention 
only to the latter ones, the interaction 
between these subsystems can be reduced to that  
of two electric dipoles 
\begin{eqnarray} 
V_{AB} = \frac{ r_i \xi_j }{ R^3 }  
\left( \delta_{ij} - 3 \frac{ R_i R_j }{ R^2 }\right),  
\label{e1}
\end{eqnarray} 
where $\delta_{ij}$ is $1$ for $i=j$ and $0$ otherwise 
and a summation over the repeated indices is implied. 
Atomic units (a.u.) are used throughout unless otherwise stated.

In the process under consideration 
one has essentially three different basic 
two-electron configurations: (I) $\psi_{\bf p}$ -- 
the incident electron is 
in the continuum, while the atomic electron 
is in the ground state; (II) $\psi_a$ --  
the incident electron is in a bound 
(ground) state and the atomic electron is excited; 
(III) $\psi_c$ -- both electrons are 
in the corresponding ground states.  

Besides the electrons, the quantum degrees of freedom 
in the process are also represented by the radiation field.  
The latter is initially in its vacuum state $\left| 0 \right\rangle$ 
and then undergoes a transition into a state 
$\left| {\bf k}, \lambda  \right\rangle$ corresponding to 
the emission of a photon with momentum ${\bf k}$ 
and polarization vector ${\bf e}_{\lambda}$ 
(with ${\bf e}_{\lambda} \cdot {\bf k} = 0$, $\lambda=1,2$).

Taking all this into account, the state vector of the system -- 
consisting of the radiation field and the two electrons, interacting with 
the nuclei, each other, and the radiation field --
can be written as 
\begin{eqnarray} 
\left| \Psi \right\rangle = %  (t)  
\left( \int d^3 {\bf p}\, b_{\bf p} \psi_{\bf p} + %  
a \psi_a \right) \left| 0 \right\rangle
+ \sum_{{\bf k}, \lambda} c_{{\bf k}, \lambda} \psi_c 
\left| {\bf k}, \lambda  \right\rangle, 
\label{e2} 
\end{eqnarray} 
where the unknown time-dependent coefficients  
$ b_{\bf p}$, $ a $ and $ c_{ {\bf k}, \lambda } $ 
satisfy the initial conditions 
$ a(t \to - \infty)=0 $, 
$c_{{\bf k}, \lambda}(t \to - \infty)=0$
and $ b_{\bf p}(t \to - \infty) = 
\delta({\bf p} - {\bf p}_i)$, where 
${\bf p}_i$ is the asymptotic 
momentum of the incident electron.  

The total (effective) Hamiltonian reads 
\begin{eqnarray}  
\hat{ H } = \hat{ H }_A + \hat{ H }_B + \hat{ H }_{ph} + 
V_{AB} + \hat{ W }.    
\label{e3} 
\end{eqnarray} 
Here, $\hat{ H }_A$, $\hat{ H }_B$ and $\hat{ H }_{ph}$ 
are the Hamiltonians for the subsystems $A$, $B$ 
and the radiation field, respectively, and 
the interaction $V_{AB}$ is given by Eq.(\ref{e1}). 
Further, 
$\hat{ W } = \frac{1}{c} \hat{ {\bf A} }( {\bf r}_1 ) \cdot \hat{{\bf p}}_1
+ \frac{1}{c} \hat{ {\bf A} }( {\bf r}_2) \cdot \hat{{\bf p}}_2$,
is the interaction of the electrons with the radiation field,
where $\hat{{\bf p}}_1$ and $\hat{{\bf p}}_2$ are the electron momenta 
and $\hat{ {\bf A} }({\bf r})$ is the vector potential for the quantized radiation field. 

Taking into account Eqs.(\ref{e1})-(\ref{e3}) one can show 
that the recombination cross section reads 
\begin{eqnarray}  
\sigma  &=& \frac{4 \pi^2  \omega_0 }{ c^3 v_i} % 
\sum_{\lambda=1}^2  
\int d \Omega_{\bf k} \bigg| 
\left\langle \phi_b \left| %
{\bf e}_{\lambda} \cdot \hat{{\bf p}}_1 %
\right| \phi_{{\bf p}_i} \right\rangle 
\nonumber \\ 
&& % 
\left. + \frac{V_{a,{\bf p}_i}}%
{ E_{p_i} - E_a + i \Gamma/2 } 
\bigg( 
\left\langle \varphi_b \left| %
{\bf e}_{\lambda} \cdot \hat{{\bf p}}_2 %
\right| \varphi_e \right\rangle  \right. 
\nonumber \\ 
&& \left.  
+ \int d^3{\bf p} 
\frac{ V_{{\bf p},a}         
\left\langle \phi_b \left| %
{\bf e}_{\lambda} \cdot \hat{{\bf p}}_1 %
\right| \phi_{\bf p} \right\rangle}%
{ \varepsilon_{p_i} - \varepsilon_p + i0 } 
\right) \bigg|^2.    
\label{e6} 
\end{eqnarray} 
Here, $\Omega_{\bf k}$ is the solid angle of the emitted photon,
$v_i$ is the incident electron velocity, 
$\phi_{\bf p}$ and $\phi_b$ are the continuum and bound states 
of this electron in the field of the nucleus $A$, 
$\varphi_b$ and $\varphi_e$ are the ground and 
excited states of the electron in the atom $B$. 
The energies of these states are denoted by 
$\varepsilon_p$, $\varepsilon_b$, 
$\epsilon_b$ and $\epsilon_e$, respectively, 
and $E_p = \varepsilon_p + \epsilon_b $ and 
$E_a = \varepsilon_b + \epsilon_e$.    
Further, 
\begin{eqnarray} 
V_{a,{\bf p}} &=& 
\left\langle \psi_a \left| V_{AB} \right| 
\psi_{\bf p} \right\rangle  
\nonumber \\ 
& = & 
\frac{ \left\langle \phi_b \left| r_i \right| 
\phi_{\bf p}  \right\rangle \, \,   
\left\langle \varphi_e \left|  
\xi_j \right| 
\varphi_b \right\rangle }{ R^3 }  
\left( \delta_{ij} - 3 \frac{ R_i R_j }{ R^2 }\right)   
\label{e7}
\end{eqnarray} 
and $\Gamma=\Gamma_a+\Gamma_{\rm rad}^{(B)}$ denotes the total width of 
the resonant two-electron state, where
$\Gamma_a = 2 \pi \left|{\bf p}_i \right| \int d\Omega_{{\bf p}_i} 
\left| V_{a,{\bf p}_i} \right|^2$ and $\Gamma_{\rm rad}^{(B)}$
are the respective contributions due to two-center Auger decay and spontaneous radiative decay of the excited state of the atom $B$. 

According to Eq.(\ref{e6}) there 
are three qualitatively different quantum
pathways for capturing the incident electron. 
(i) The transition 
$\phi_{{\bf p}_i} \to \phi_b$ occurs 
without the participation of the atom $B$  
via direct photon emission. % (the transition matrix element in the 
% the first line of Eq.(\ref{e6})). 
(ii) The electron is captured from the state  
$\phi_{{\bf p}_i}$ into the state $\phi_b$ 
by inducing the transition $\varphi_b \to \varphi_e$ 
in the atom $B$; the latter afterwards deexcites 
by photon emission. % (the transition matrix elements in the 
%the second line of Eq.(\ref{e6})). 
(iii) The incident electron undergoes the transitions  
$\phi_{{\bf p}_i} \to \phi_b \to \phi_{\bf p} \to \phi_b$ 
in which the first two steps are accompanied by 
the (radiationless) transitions $\varphi_b \to \varphi_e \to \varphi_b $ 
in the atom $B$ while the last one proceeds 
via photon emission. 
The pathways (ii) and (iii) are resonant and become efficient only if 
the energies $E_{p_i}$ and $E_a$ are very close.

When considering solely the pathway (ii) in Eq.\,\eqref{e6}, we obtain the partial cross section for 2CDR. Averaged over incoming electron angles, it reads
\begin{eqnarray}
\sigma_{\rm 2CDR} = \frac{\pi}{p_i^2}
\frac{\Gamma_a \Gamma_{\rm rad}^{(B)}}{(E_{p_i}-E_a)^2+\Gamma^2/4}\,.
\label{2CDR}
\end{eqnarray}
We note that, when more than one intermediate state of the same energy exists in atom $B$ (reachable by a dipole transition), the cross sections \eqref{e6} and \eqref{2CDR} involve an appropriate sum over these states.

Let us now consider two illustrative examples. First we discuss a very simple and basic situation which can be treated easily in detail. Assume that an electron recombines with a proton into the $1s$ state while a He$^+$ ion is located nearby which may be excited to a $2p$ level with magnetic quantum number $m$: $e^- + {\rm H}^+ + {\rm He}^+(1s) \to {\rm H}(1s) + {\rm He}^+(2p_m) \to {\rm H}(1s) + {\rm He}^+(1s) + \gamma$.
In this case the electron wave functions are known analytically and we obtain
\begin{eqnarray}
\Gamma_a(2p_m) = \frac{2^{21}\pi}{3^{11}}\frac{C_m}{R^6} \frac{Z_A^6}{Z_B^2\omega_0^5}\frac{e^{-(4Z_A/p_i)
\arctan(p_i/Z_A)}}{1-e^{-2\pi Z_A/p_i}}\,,
\end{eqnarray}
with $C_0=1$ and $C_{\pm 1}=1/4$ for the intermediate $2p_m$ states in He$^+$ with $m=0$ and $m=\pm 1$, respectively. The quantization axis is chosen along the internuclear separation vector ${\bf R}$. For the nuclear charges $Z_A=1$ and $Z_B=2$ of the example, we obtain $\Gamma_a(2p_0)=4\Gamma_a(2p_{\pm 1})\approx 0.08/R^6$. The radiative decay width  $\Gamma_{\rm rad}^{(B)}=2^{17}\omega_0^3/(3^{11}c^3Z_B^2)\approx 2.4\times 10^{-7}$\,a.u. \cite{Bethe} surpasses the Auger widths for $R\gtrsim 8$\,a.u. Assuming a resonant electron energy $\varepsilon_{p_i}=1$\,a.u., the 2CDR cross section \eqref{2CDR} largely exceeds single-center RR up to $R\approx 100$\,a.u. (see Fig.\,2).
Therefore, even taking into account that only a small fraction of the electrons might be able to participate in 2CDR because of its resonant character, it is obvious that this process can compete with and even strongly dominate over single-center RR. 

\begin{figure}[hb]
\begin{center}
\resizebox{6.5cm}{!}{\includegraphics{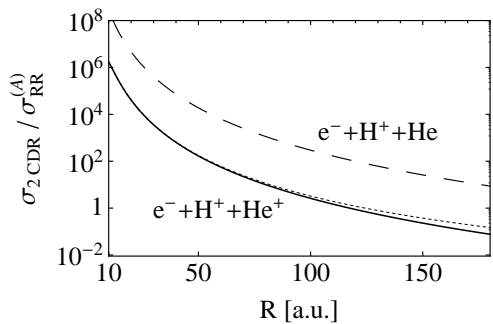}}
\vspace{-0.3cm}
\caption{Ratio of $\sigma_{\rm 2CDR}$ to $\sigma_{\rm RR}^{(A)}$ as a function of internuclear distance. The solid and dotted lines refer to the system '$e^-$ + H$^+$ + He$^+$($1s$)', with the dotted line including retardation effects. For comparison, the system '$e^-$ + H$^+$ + He($1s^2$)' is described by the dashed line. In both cases, the change of the incident electron energy is assumed to be resonant with the corresponding first dipole-allowed transition energy for the center $B$ ($\varepsilon_{p_i}=1$\,a.u. and 0.28\,a.u., respectively).}
\end{center} 
\end{figure}

In a complete picture of the recombination process, all the channels and their interference must be accounted for since they lead to the same final state [see Eq.\,\eqref{e6}]. As a result, we obtain
\begin{eqnarray}
\sigma = \frac{\sigma_{\rm RR}^{(A)}}{3} \sum_{m=-1}^1
\frac{[E_{p_i}-E_a+q_m\Gamma_a(2p_m)]^2+(\Gamma_{\rm rad}^{(B)}/2)^2}{(E_{p_i}-E_a)^2+\Gamma^2/4}.
\end{eqnarray}
Here, $q_0=-2 \left| \left\langle \varphi_b \left| 
\mbox{\boldmath{$\xi$}}\!\cdot\!{\bf R}\right| \varphi_e \right\rangle \right|^2\! /(R^5 \Gamma_a(2p_0))$
and $q_{\pm 1}=-2q_0$ can be termed as two-center Fano parameters. Similarly to the usual Fano parameters in the case of single-center DR \cite{AMueller,Fano}, they describe the relative strength of the indirect (2CDR) versus the direct (RR) capture channels as a function of the incident electron energy. However, in contrast to single-center DR, in our case the Fano parameters depend on the internuclear distance, implying that the shape of the recombination cross section varies with $R$ (see Fig.\,3 for an illustration).

\begin{figure}[h]
\begin{center}
\includegraphics[height=4.2cm]{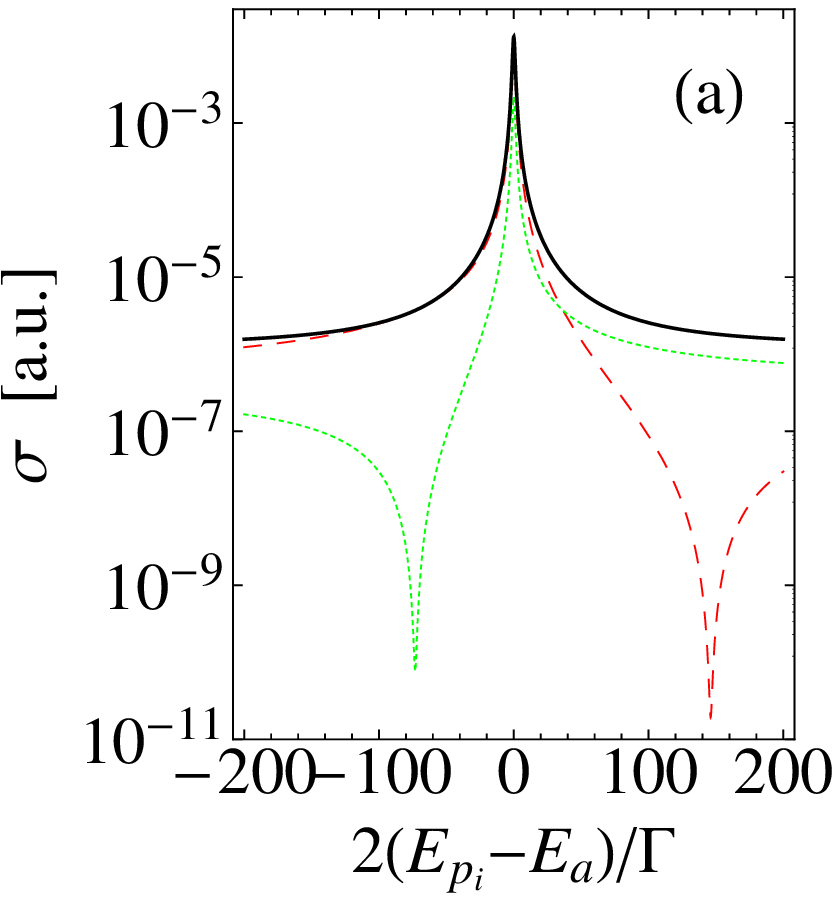}
\includegraphics[height=4.2cm]{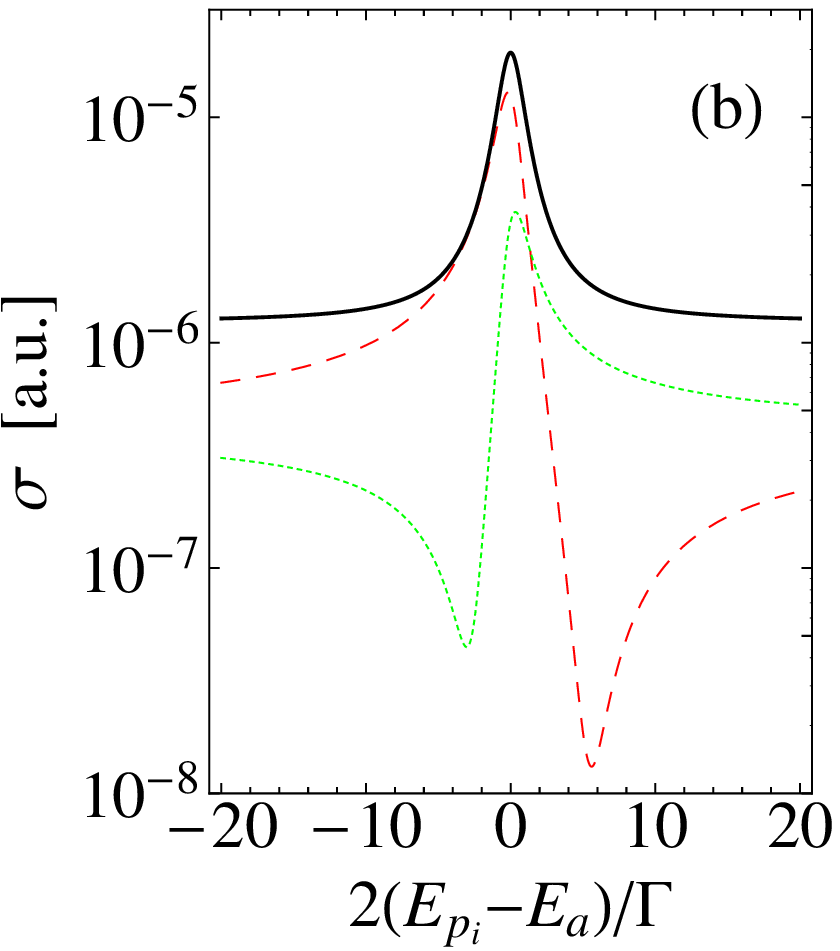}
\vspace{-0.3cm}
\caption{Fano profiles for the $2p_0$ (red dashed curves) and $2p_1$ (green dotted curves)
intermediate states and for the whole process $e^- + {\rm H}^+ + {\rm He}^+(1s) \to {\rm H}(1s) + {\rm He}^+(1s) + \gamma$ (solid curves). The internuclear distances are (a) $R=25$\,a.u. and (b) $R=75$\,a.u.}
\end{center} 
\end{figure}

2CDR in systems similar to the previous example is of relevance in plasma-like environments such as warm dense matter \cite{WDM} where typical densities $\sim 0.1$\,g/cm$^3$ and temperatures $T$ corresponding to $kT\sim 1$\,eV prevail. This state of matter is found, e.g., in astrophysical objects or thermonuclear fusion plasmas and can be produced in the laboratory by intense laser-solid interaction \cite{WDMexp}. 2CDR can clearly affect the time evolution and charge balance of warm dense matter systems. Indeed, at $\omega_0\sim 1$\,a.u. and $R\sim 10$\,a.u., it dominates over RR by $\approx 7$ orders of magnitude at the resonance. For $\Gamma\sim 10^{-7}$\,a.u., a fraction of $\Gamma/kT\sim 10^{-6}$ of all electrons contributes to 2CDR implying that it greatly exceeds RR also in the total number of recombination events.

Within a more general (but less detailed) approach, 2CDR can also be treated to a good approximation in much more complex systems. To this end, we note that the two-center Auger width can be expressed in terms of $\Gamma_{\rm rad}^{(B)}$ and the photoionization cross section $\sigma_{\rm PI}^{(A)}$ of atom $A$ as
$\Gamma_a \sim (1/R^6)(c/\omega_0)^4\sigma_{\rm PI}^{(A)}\Gamma_{\rm rad}^{(B)}$
up to a numerical prefactor of order unity \cite{surface}.
The photoionization cross section is related to the RR cross section $\sigma_{\rm RR}^{(A)}$ via the principle of detailed balance: $p_i^2  \sigma_{\rm RR}^{(A)} \propto (\omega_0/c)^2 \sigma_{\rm PI}^{(A)}$. Exactly on the resonance, the ratio between the cross sections for 2CDR and single-center RR thus becomes
\begin{eqnarray}
\frac{\sigma_{\rm 2CDR}}{\sigma_{\rm RR}^{(A)}} \sim \left(\frac{c}{R \omega_0}\right)^6,
\label{ratio}
\end{eqnarray}
assuming that $\Gamma_a\lesssim\Gamma_{\rm rad}^{(B)}$ holds which will be the case at sufficiently large values of $R$.

The next example illustrates that 2CDR can be of relevance in (bio)chemical environments as well, where free electrons are always present due to photoionization by ultraviolet radiation \cite{free_electrons}. When an alkali salt such as NaCl is dissolved in water, the molecule dissociates into an alkali cation (Na$^+$) and a remaining anion (Cl$^-$) which both are surrounded by water molecules forming hydration shells. We consider the Na$^+$(H$_2$O)$_n$ complex where on average $n=6$ water molecules shield the cation at a mean distance of about $R\approx 6$\,a.u. \cite{hydration}. The ionization energy of neutral Na is 5.14\,eV and the first (photo) absorption band in water begins at $\approx 6.7$\,eV reaching its maximum at $\approx 7.5$\,eV. 
According to Eq.\,\eqref{ratio}, the assistance by a single water molecule at $R=10$\,a.u. and resonant electron energies $\varepsilon_{p_i}\approx 1.5$--4\,eV would enhance the recombination with the Na$^+$ ion by $\approx 10$ orders of magnitude.
This number can serve as a lower estimate for the enhancement effect at the real value of $R\approx 6$\,a.u. (where the dipole approximation might be already not very reliable).
The dramatic enhancement is further amplified by the fact that more than one water molecule surrounds the cation, which can be taken into account approximately by multiplication with the coordination number $n=6$ \cite{ICEC}. Since the absorption band in water is relatively broad, we may conclude that there is an enormous effect also for the total number of $e^-$ + Na$^+$ recombination events (integrated over the incident electron energies).

This example clearly demonstrates that 2CDR can be important in chemical systems. It may be generalized to solvated biomolecules where the enhanced recombination can lead to accelerated bond breaking and dissociation, which may initiate a subsequent reaction chain producing chemically active fragments and radicals \cite{free_electrons,Hvelplund}.

Note that dedicated experiments on hydrated atoms and molecules have become possible just recently utilizing liquid jets of aqueous solution injected into vacuum chambers. Atomic collision studies \cite{Hvelplund} and ultrafast photoelectron spectroscopy \cite{jets} on solvated substances are carried out this way, for example. A similar method could be applied to study water-catalyzed 2CDR.

Let us very briefly compare 2CDR with single-center DR. The latter is described by a formula analogous to Eq.\,\eqref{2CDR} but occurs, in general, at different resonance energies. Single-center Auger rates $\Gamma_a^{(A)}$ are typically orders of magnitude larger than the radiative ones. As a result, at comparable resonance energies, the ratio
$\sigma_{\rm 2CDR}/\sigma_{\rm DR}^{(A)}\propto \Gamma_a\Gamma_a^{(A)}/(\Gamma_{\rm rad}^{(A)}\Gamma_{\rm rad}^{(B)})$ can largely exceed unity.  One can also show that the ratio of the energy-integrated cross sections (i.e. of the resonance strengths) varies in a broad range depending on the system parameters and can be both smaller and larger than unity. The latter is the case, for instance, in the example of Na$^+$ in water considered above. Such competitiveness of 2CDR with single-center DR (despite $\Gamma_a^{(A)}\gg \Gamma_a$) is due to the fact that here the majority of initially captured electrons remains bound because $\Gamma_a\lesssim\Gamma_{\rm rad}^{(B)}$.

Finally, we would like to mention that 2CDR (as well as the process considered in \cite{ICEC}) can be viewed as a kind of three-body recombination in which -- in contrast to its standard case -- the role of the free assisting electron is played by an (initially) bound electron.

In conclusion, we have shown that the presence of neighboring atomic centers at nanometer distances can resonantly enhance the recombination (or attachment) probability by orders of magnitude. The resonance effect is so strong that it can largely outperform RR even after averaging over incident electron energy distributions much broader than the resonance width. Representing a hitherto unexplored and efficient recombination mechanism, this two-center process is of general interest by itself and deserves further study. Besides, it may play a significant role in chemical and dense plasma environments where it can substantially affect the quantum dynamics and time evolution of the system.

We thank J. Evers, R. Moshammer, B. Najjari, A. Senftleben, and A. Wolf for useful conversations. This work was supported by Helmholtz Alliance HA216/EMMI.

\end{document}